\documentclass[aps,pra,nofootinbib,superscriptaddress,longbibliography,twocolumn,floatfix,10pt]{revtex4-1}
\usepackage{graphicx}
\usepackage{amsmath,amssymb,amsfonts}
\usepackage{textcomp,keyval,wrapfig,float,mathtools,amsmath,amsfonts,amssymb,url,physics,subcaption,gensymb,float,acro,braket}
\usepackage{dcolumn}
\usepackage[version=4]{mhchem}
\usepackage[linkcolor=black]{hyperref}
\usepackage[capitalize]{cleveref}
\usepackage{diagbox}
\usepackage{orcidlink}
\usepackage[sort&compress]{natbib}

\usepackage[latin1]{inputenc}
\usepackage{amsthm}

\usepackage{psfrag}
\usepackage{upgreek}
\usepackage{bbold}
\usepackage{verbatim}
\usepackage{cancel}
\usepackage{leftidx}
\usepackage{mathtools}
\usepackage{esint} 
\usepackage[justification=justified]{caption} 
\usepackage[dvipsnames]{xcolor}
\usepackage{dsfont}
\usepackage{xfrac}
\usepackage[normalem]{ulem}   

\usepackage{braket}
\usepackage{ragged2e}

\renewcommand{\vec}[1]{\mathbf{#1}}

\usepackage{color} 

\begin{document}

\hypersetup{pdftitle={A simple
quantum dot: Numerical and variational solutions}}
\title{A simple
quantum dot: Numerical and variational solutions}
\author{Connor M. Walsh\,\orcidlink{0009-0002-6311-9108}}
 \email{cmwalsh@ualberta.ca}
 \affiliation{Department of Physics, University of Alberta, Edmonton, AB, Canada T6G~2E1}
\author{Ian MacPherson\,\orcidlink{0000-0002-8401-5458}}
 \affiliation{Department of Physics, University of Alberta, Edmonton, AB, Canada T6G~2E1}
\author{Davidson Noby Joseph\,\orcidlink{0009-0008-2420-0951}}
 \affiliation{Department of Physics, University of Alberta, Edmonton, AB, Canada T6G~2E1}
\author{Suyash Kabra\,\orcidlink{0009-0004-9321-846X}}
 \affiliation{Department of Physics, University of Alberta, Edmonton, AB, Canada T6G~2E1}
\author{Ripanjeet Singh Toor\,\orcidlink{0009-0006-8987-0724}}
 \affiliation{Department of Physics, University of Alberta, Edmonton, AB, Canada T6G~2E1}
\author{Mason Protter\,\orcidlink{0000-0001-8942-4448}}
 \affiliation{Department of Physics, University of Alberta, Edmonton, AB, Canada T6G~2E1}
\author{Frank Marsiglio\,\orcidlink{0000-0003-0842-8645}}
 \email{fm3@ualberta.ca}
 \affiliation{Department of Physics, University of Alberta, Edmonton, AB, Canada T6G~2E1}
 \affiliation{Theoretical Physics Institute \& Quantum Horizons Alberta, University of Alberta, Edmonton, Alberta T6G 2E1, Canada}

\begin{abstract}
We describe a simple quantum dot that consists of two crossed two-dimensional troughs. As such there is no potential well; nonetheless, this geometry gives rise to a bound state, centred on the point at which these troughs cross one another. This problem is interesting both because the existence of a bound state may surprise students and because it can be solved using a variety of computational techniques, including matrix mechanics, finite differences, and mode matching. We present these methods and show how the mode-matching method in this case provides the most accurate solution to the problem. Additionally, the mode-matching method can be used to generate a simple wave function that yields the lowest energy known to date to arise out of an analytical variational solution for this problem.
\end{abstract}

\maketitle

\section{Introduction}\label{intro}

\begin{figure}[tb]
    \centering
    \includegraphics[width=1.0\linewidth]{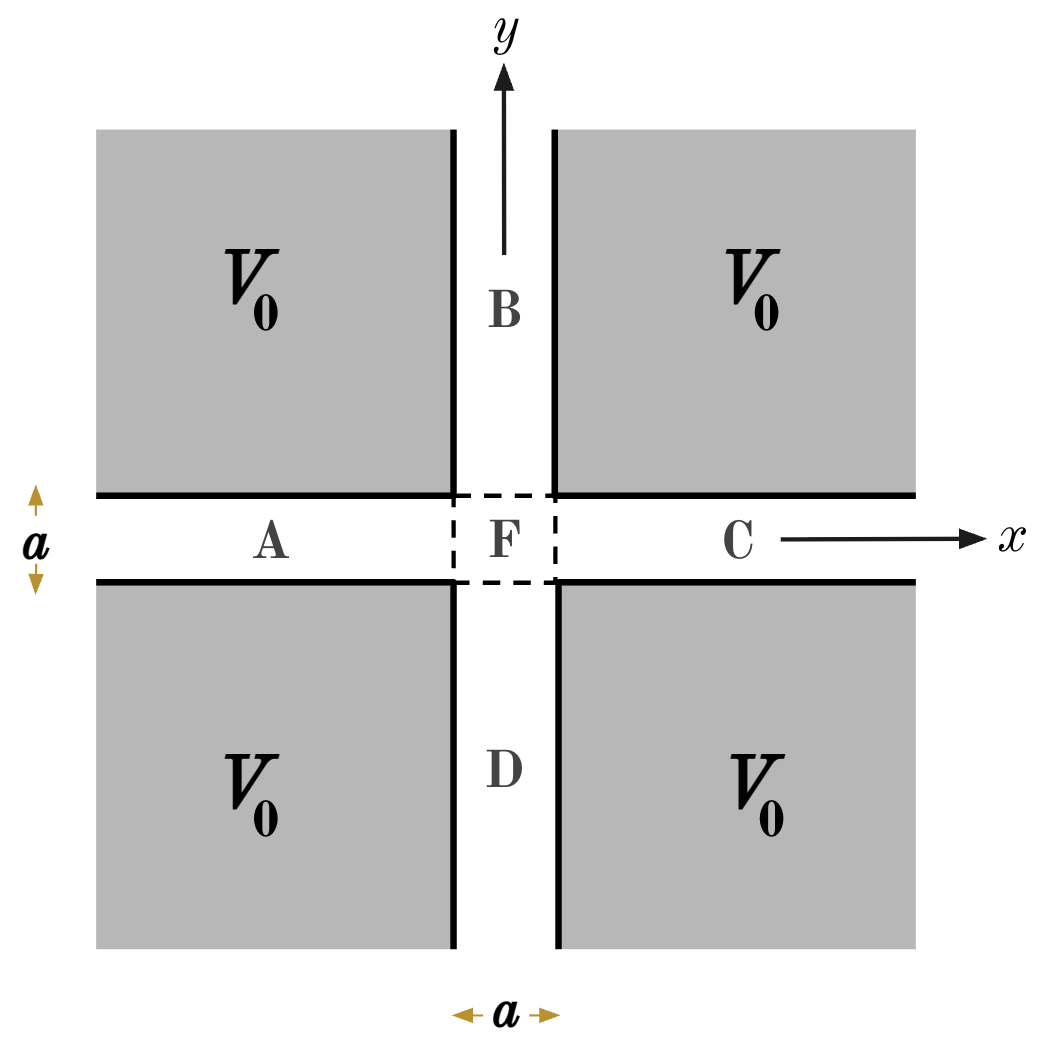}
    \caption{\justifying
    Schematic of the crossed troughs of width $a$ (white regions). The grey-shaded regions have a very large or infinite potential $V_0$ that confines the particle to the troughs. Regions labelled by the letters A through F are relevant for the mode-matching method described in \cref{modematching}. No bound state is expected to exist, classically.}
    \label{fig1}
\end{figure}

This paper proposes a compelling problem for an undergraduate quantum mechanics course, extending the ubiquitous problem of a particle in a potential well. We consider the two-dimensional system shown in Fig.~\ref{fig1}, consisting of two infinitely long perpendicular troughs of zero potential that separate regions of very large or infinite potential. Somewhat surprisingly, while classically no bound state exists, the quantum mechanical problem has a bound state, as was shown using the variational principle in Griffiths's Introduction to Quantum Mechanics \cite{griffiths18}.

Students will already be familiar with the related one-dimensional problem of a particle of mass $m_0$ in a potential well of width $a$. When $V_0$ is infinite, the ground-state energy is
\begin{equation}
    E_{\rm t} = \frac{\hbar^2\pi^2}{2m_0a^2}. \label{Et}
\end{equation}
We include the subscript ``${\rm t}$'' which stands for ``threshold'' because a particle freely propagating along one arm of the cross in \cref{fig1} is necessarily bound in the transverse direction and thus has an energy at least equal to that of a particle in a one-dimensional square well with barrier height $V_0$. As $V_0 \rightarrow \infty$, this minimum energy is given by $E_{\rm t}$; for the geometry in \cref{fig1}, if a solution exists with energy lower than $E_{\rm t}$, such a state is necessarily bound. Additionally, $E_{\rm t}$ sets a scale for the energy in this problem, allowing us to define a dimensionless energy $\epsilon \equiv E/E_{\rm t}$.

We wish to adopt the crossed-trough example as a case study of a problem that is interesting (classically, there is no bound state) and yet is amenable to straightforward numerical solution in a variety of ways. A direct solution to the Schr\"odinger equation is not presented in Griffiths, and that is what we wish to do in the present paper, in a form suitable for undergraduates with access to a diagonalization subroutine, as is available on any computing platform. We will discuss three numerical methods, each with its distinct advantages: (i) matrix mechanics, (ii) finite differences, and (iii) mode matching.

The outline of this paper is as follows. In Section~\ref{matrixmechanics}, we present the solution using matrix mechanics. We consider this the most straightforward and flexible method.
While matrix mechanics is rare in elementary textbooks, the technique has been applied to the one-dimensional harmonic oscillator \cite{marsiglio09} and to one-dimensional \cite{pavelich15} and two-dimensional \cite{pavelich16} periodic array problems in this journal.

In Section~\ref{finitedifferences}, we briefly outline the method of finite differences. This methodology is often used to discretize a differential equation. 
A more sophisticated version of this method, called the relaxation method, was used by Schult et al. \cite{schult89} to solve the problem at hand
(see details of the method in Ref.~[\onlinecite{press86}]).
Londergan and Murdock \cite{londergan12} also used the relaxation method for a similar model. We will present a simplified version of this finite-difference method (relaxation is not needed to solve our problem).

In Section~\ref{modematching}, we present the least known and most accurate method for this problem, the so-called mode-matching method. To our knowledge, this method was first discussed in Ref.~[\onlinecite{schult89}] and in the context of isospectral cavities in Ref~[\onlinecite{wu95}], and then for related problems in Ref.~[\onlinecite{londergan12}]. This method converges very quickly and is also easy to implement. For these reasons we devote a little more attention to this method, both in the main text and in the Appendix.

In Section~\ref{tightbinding}, we examine the simplified problem of crossed one-dimensional atomic chains, described by a tight-binding Hamiltonian, to illustrate that the physics of a bound state emerges even for this effective Hamiltonian. One can readily compute the energy levels for this problem, revealing that the ground state is indeed a bound state. Thus, this effective model (tight-binding) qualitatively describes
the creation of a bound ground state.

Finally, in Section~\ref{sec_var}, we close with a simple analytical variational solution that arises from the simplest approximation to the mode-matching method. This very simple wave function outperforms (i.e. has a lower energy than) any other known
analytical variational solutions.

Although we strive to keep each section self-contained, we have relegated many of the details to the appendices. In this way, we hope the reader can easily follow the main text. At the same time, for the reader interested in one particular technique, the appendices and references should provide sufficient guidance. In addition to the novelty of a quantum bound state where no classically bound state exists, we hope the reader will be motivated to further learn one of the techniques described here and apply it to a new problem. Moreover,  Section~\ref{sec_var} shows how inspired analytical results can emerge from the consideration of numerical results.

\section{Matrix Mechanics}
\label{matrixmechanics}

We wish to solve the time-independent Schr\"odinger equation,
\begin{equation}
        -\frac{\hbar^2}{2m_0} \nabla^2 \Psi(x,y) +V(x,y)\Psi(x,y) = E\Psi(x,y), \label{SEwithV} 
\end{equation}
for the potential depicted in \cref{fig1}. However, to use matrix mechanics, it is necessary to do two things: (i) use $V_0 < \infty$, and (ii) enclose the potential in a two-dimensional infinite square well, i.e. a box of side length $L$, as depicted in \cref{appa_fig1} in Appendix~\ref{matrixappendix}. Analytically, the potential is now given by
\begin{equation}
    V(x,y) = \begin{cases}
        0 & \text{for } |x| < \frac{a}{2} \text{ and } |y| < \frac{L}{2}, \\
         & \text{ or } |y| < \frac{a}{2} \text{ and } |x| < \frac{L}{2}, \\
        V_0 & \text{for }\frac{a}{2} < |x| < \frac{L}{2} \text{ and } \frac{a}{2} < |y| < \frac{L}{2},
         \\
        \infty & \text{otherwise},
    \end{cases} \label{V0}
\end{equation}
where $V_0\gg E_{\rm t}$ should be large compared to any other energy scale in the problem. We need $L \gg a$ for the box to have no influence on the bound state; we find $L = 5a$ suffices. Here, the reason for the confinement to a two-dimensional box is that it allows us to construct normalizable basis states \cite{marsiglio09}.
 
Following Ref.~[\onlinecite{marsiglio09}], we expand the unknown wave function in terms of the known eigenstates of the two-dimensional infinite square well of side length $L$,
\begin{equation}
    \Psi(x,y) = \sum_{m_x=1}^\infty\sum_{m_y=1}^\infty c_{m_x,m_y} \phi_{m_x,m_y}(x,y), \label{MMexpansion}
\end{equation}
where
\small
\begin{equation}
    \phi_{m_x,m_y}(x,y)\equiv \sqrt{\frac{2}{L}} \sin\biggl[\frac{m_x\pi}{L} \Bigl(x + \frac{L}{2}\Bigr)\biggr] \sqrt{\frac{2}{L}}\sin\biggl[\frac{m_y\pi}{L} \Bigl(y + \frac{L}{2}\Bigr)\biggr]. \label{boxbasis}
\end{equation}
\normalsize

We rewrite the Schr\"odinger equation (\ref{SEwithV}) with this basis, and then take the inner product with an arbitrary basis state (see Appendix~\ref{matrixappendix} and also Ref.~[\onlinecite{pavelich16}]). This procedure results in a simple dimensionless matrix equation,
\begin{equation}
\sum_{m=1}^\infty h_{nm} c_m = \epsilon c_n,  \label{mat_eq}
\end{equation}
where $m \equiv (m_x,m_y)$ and similarly for $n$. The definition of $h_{nm}$ and other details of the derivation and truncation procedures are provided in \cref{matrixappendix}. Diagonalization of $h_{nm}$ provides eigenvalues $\epsilon\equiv E/E_{\rm t}$ and eigenvector solutions for a given choice of $L$, $a$, and $V_0$. 

\begin{figure}[tb]
    \centering
    \includegraphics[width=\linewidth]{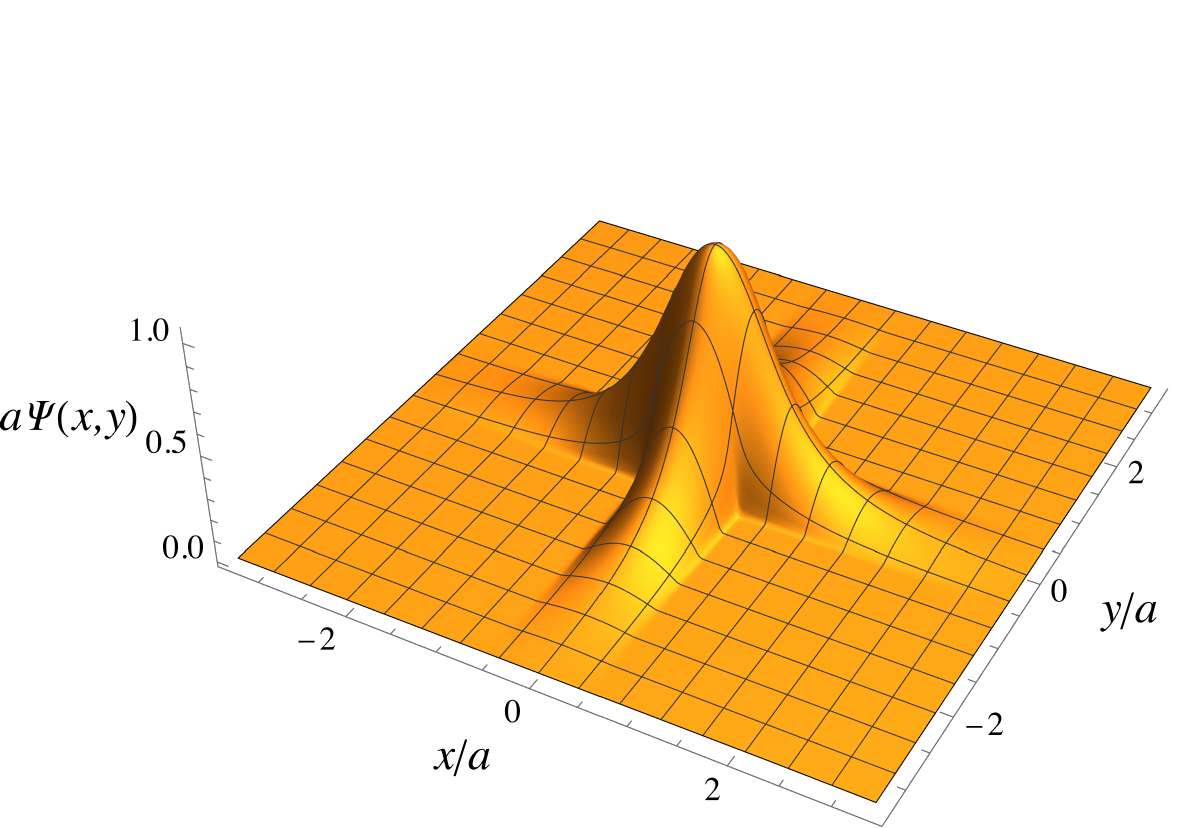}
    \caption{\justifying
    Ground state wave function, computed with Eq.~\ref{MMexpansion} for $L/a=20/3$ and ${v}_0 \equiv V_0/E_{\rm t} = 500$. Note that the wave function is localized near the origin. We have confirmed that similar plots are obtained as $L/a$ is increased and for a large range of $V_0$ values.}
    \label{fig2}
\end{figure}

Figure~\ref{fig2} shows the ground state wave function as calculated through the procedure just described, with $L/a=20/3$ and $v_0 \equiv V_0/E_{\rm t} = 500.$ As expected, the ground state is bounded near the origin where the two troughs intersect. The wave function has essentially already decayed to zero at the boundaries ($x,y = \pm L/2$). Indeed, we have confirmed that the wave function remains unchanged for larger values of $L/a$. This shows that the box edges at $\pm L/2$ play no role in the results, as should be the case.

\begin{figure}[tb]
    \centering
    \includegraphics[width=1.0\linewidth]{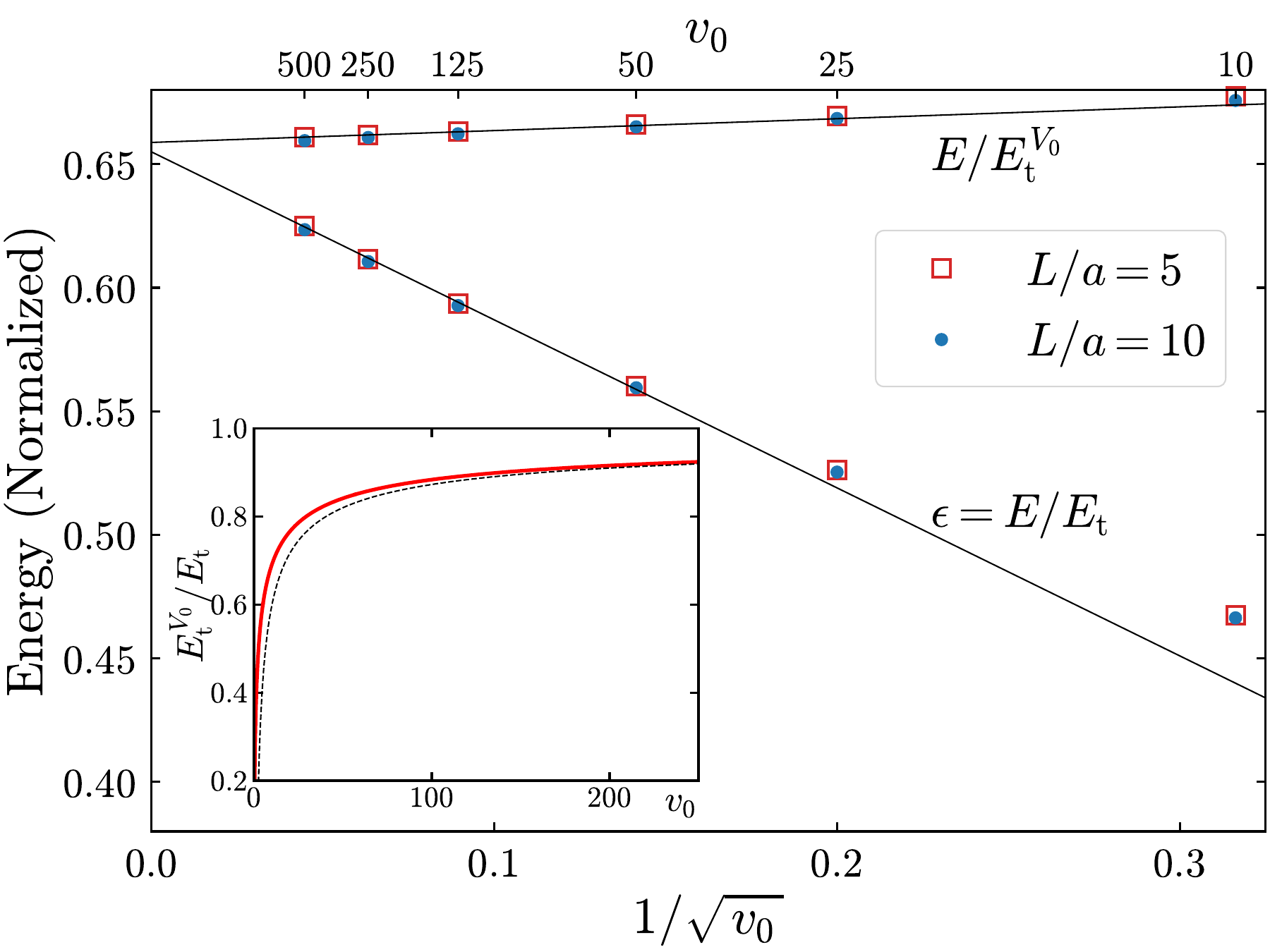}
    \caption{\justifying
    Bound state energy versus $1/\sqrt{{v}_0}$ for $L/a=5$ (unfilled square symbols) and for $L/a=10$ (filled circular symbols). The actual value of ${v}_0 \equiv V_0/E_{\rm t}$ corresponding to each symbol is shown across the top of the graph. Note that the lower set of symbols shows $\epsilon \equiv E/E_{\rm t}$, while the upper set of symbols shows $E/E_{\rm t}^{V_0}$, which is a better measure of boundedness. In view of this, the relatively shallow slope of the upper curve shows that the level of boundedness varies less with $V_0$ than the lower set of points might have implied. Note also that the two sets of symbols (circles and squares) lie nearly on top of one another, showing that for these values of $L/a$, the energy hardly depends on the choice of $L$. The inset (solid red curve) shows the normalized threshold energy $E_{\rm t}^{V_0}/E_{\rm t}$ for a bound state as a function of ${v}_0$. The dashed black curve gives the first-order analytical expansion (see Eq.~(\ref{expansion}) in Appendix~\ref{matrixappendix}).}
    \label{fig3}
\end{figure}

As an aside, we can ask the question: How universal is the existence of a bound state due to the troughs? We can answer this, thanks to the ability to vary $V_0$ with this method. In Fig.~\ref{fig3} (main panel, lower set of points) we show the ground state energy, normalized to $E_{\rm t}$, as a function of $1/\sqrt{{v}_0}$. Numerical results for a few representative values of ${v}_0$ are given by the symbols. First, note that there are only small differences between the results for $L/a=5$ and those for $L/a=10$, indicating that the system is sufficiently large that the boundary effects due to the artificial box are negligible. Secondly, note that as $V_0$ increases (moving towards the origin in this graph), the energy $E/E_{\rm t}$ also increases. This is expected since the particle is becoming increasingly confined to the troughs, and therefore its energy is increasing as $V_0$ increases. On the other hand, the only reason a bound state exists at all is due to the troughs, which become increasingly well defined as $V_0$ increases, suggesting that the state should become more bound (i.e. its energy should decrease) as $V_0$ increases. This latter conclusion is not supported by the lower set of points in the main panel of Fig.~\ref{fig3}.

The resolution of this apparent contradiction is that the criterion for boundedness is itself a function of $V_0$, as described in Appendix~\ref{matrixappendix}. For finite $V_0$, the relevant threshold energy is actually not $E_{\rm t}$, which corresponds to the ground state energy of an infinite square well, but the ground state energy of a finite square well (with strength $V_0 < \infty$); we denote this threshold energy by $E_{\rm t}^{V_0}$. For the upper set of points in Fig.~\ref{fig3}, we normalize the ground state energy to this new threshold, $E_{\rm t}^{V_0}$, and now these values decrease as $V_0$ increases, indicating a higher degree of boundedness as $V_0$ increases (even though the absolute ground state energy increases as $V_0$ increases). This is now as we expect, and as $V_0$ increases, the ratio $E/E_{\rm t}^{V_0}$ decreases, indicating a more bounded state. For reference, the inset of Fig.~\ref{fig3} shows how the threshold energy $E_{\rm t}^{V_0}$ varies with $V_0$.

Our main interest, though, is in the result for $V_0 \rightarrow \infty$. Linear fits to $E/E_{\rm t}^{V_0}$ and $E/E_{\rm t}$ are shown in Fig.~\ref{fig3}, and the extrapolated energy value from either fit is clearly close to $0.66$, consistent with the result quoted in Ref.~[\onlinecite{schult89}]. We could determine this value more accurately here, by using larger values of $V_0$, for example, but we will defer the determination of a more accurate result to the mode-matching method to be described in Section~\ref{modematching}. Also note that 
normalizing the bound state energy to the $V_0$-dependent threshold energy ($E_{\rm t}^{V_0}$) shows that the boundedness is less dependent on $V_0$ than is implied by the values of $\epsilon$.

In summary, the results shown in this section indicate that matrix mechanics is a relatively straightforward method to obtain bound-state wave functions and energies. This method is more flexible than the methods described in the following sections, with respect to the characteristics of the potential. For example, we have explored the possibility of non-infinite barrier potentials and found that boundedness occurs even for relatively shallow potential barriers, and that the degree of boundedness is surprisingly insensitive to the barrier height.

\section{Finite Differences}\label{finitedifferences}

Another standard procedure for obtaining numerically exact solutions to the Schr\"odinger equation is to discretize the domain by introducing a grid of points on which to define the wave function, and to approximate all derivatives using finite-difference formulas. In this section, we shift our focus to the case where $V_0\rightarrow\infty$, so that the particle is perfectly confined to the troughs. As in \cref{matrixmechanics}, the troughs are terminated at a distance of $L/2$ from the origin, in all four directions. The Schr{\"o}dinger equation now becomes
\begin{equation}
    -\frac{\hbar^2}{2m_0}\nabla^2\Psi(x,y) = E\Psi(x,y), \label{SE_no_V}
\end{equation}
defined only on the white regions of \cref{fig1}, with Dirichlet boundary conditions on the black boundaries and at the trough ends.

Substituting the discrete Laplacian (see Appendix~\ref{finteappendix})
into \cref{SE_no_V} results in a matrix equation
\begin{equation}
    \sum_i \bar{L}_{ij} \Psi_{j} = \epsilon\Psi_i,
    \label{Laplacian_matrix}
\end{equation}
which relates the unknown values of $\Psi_i\equiv\Psi(x,y)$ at the interior grid points. Details of the discretization procedure and the matrix $\bar{L}$ are provided in \cref{finteappendix}. 

Diagonalization of the sparse matrix $\bar{L}$ gives numerical approximations to the eigenvalues $\epsilon\equiv E/E_{\rm t}$ and eigenvector solutions of \cref{SE_no_V} for particular choices of $L$ and $a$; the density of internal grid points is then increased until the numerical results converge to a desired precision. As Table~\ref{finitegs} in Appendix~\ref{finteappendix} shows,
in the limit $L/a\rightarrow\infty$, this yields a ground state energy of $\epsilon=0.660$, in agreement with the other methods outlined in this paper. 

The finite-difference method yields good results with minimal effort; however the convergence rate with mesh size is slow, and the flexibility is not as good as the previous method. For example, solving the problem with finite $V_0$ is not straightforward, and it is difficult to implement on geometries with rounded edges or with non-right angles. The calculated energies approach the most accurate ground state energy provided with the mode-matching method (see Section~\ref{modematching}). Similarly, the wave functions are very accurate; we do not provide a separate 2D plot as it is indistinguishable from that presented in Section~\ref{matrixmechanics}. Small discrepancies with the matrix mechanics result exist because here $V_0 \rightarrow \infty$, and so no leakage of the wave function outside of the troughs can occur. This will be further explored in Section~\ref{modematching}.

\section{Mode-Matching Method}
\label{modematching}
Another method for tackling such problems is known as mode matching, which can be used to treat troughs of either finite or infinite length. Here, we will discuss troughs of infinite length, as shown in \cref{fig1}. For this method, it is required that $V_0\rightarrow\infty$ so that the particle is perfectly confined within the troughs. However, this method has the benefit of faster numerical convergence than the other two methods presented, and therefore yields our most accurate solution. 

We now seek solutions to the Schr{\"o}dinger equation~(\ref{SE_no_V}), with Dirichlet boundary conditions along the trough walls as well as at $\pm\infty$. The geometry is divided into regions labelled A through F in \cref{fig1}, on which \cref{SE_no_V} can be solved analytically and expressed as a series of Fourier modes. Matching the wave function and its derivatives across the boundaries between regions (denoted by dashed lines in the figure) results in a system of equations that relate the expansion coefficients to one another. Solutions to this system of equations give the energies and wave functions that solve \cref{SE_no_V}. 

To proceed, we first write exact solutions that are valid in each of the regions A through F, assuming that $E < E_{\rm t}$; these solutions are linear combinations of basis states for the different regions, as follows (see Appendix~\ref{modeappendix} for details):

\begin{align}
    \Psi_{\rm A,C}(x,y) &= \sum_{n=1}^\infty A_n e^{\lambda_n\bigl(\frac{a}{2}-|x|\bigr)} \sin \biggl(\frac{n\pi}{a}\Bigl(y + \frac{a}{2}\Bigr)\biggr), \label{psiAC} \\
    \Psi_{\rm B,D}(x,y) &= \sum_{n=1}^\infty A_n e^{\lambda_n\bigl(\frac{a}{2}-|y|\bigr)} \sin \biggl(\frac{n\pi}{a}\Bigl(x + \frac{a}{2}\Bigr)\biggr), \label{psiBD} \\
    \Psi_{\rm F}(x,y) &= \sum_{n=1}^\infty A_n \frac{\cosh\lambda_nx}{\cosh\frac{\lambda_na}{2}} \sin\biggl( \frac{n\pi}{a}\Bigl(y+\frac{a}{2}\Bigr)\biggr) \phantom{.} \notag \\
        &\,+  \sum_{n=1}^\infty A_n \frac{\cosh\lambda_ny}{\cosh\frac{\lambda_na}{2}} \sin\biggl( \frac{n\pi}{a}\Bigl(x + \frac{a}{2}\Bigr)\biggr). \label{psiEsym}
\end{align}
where $n$ is a positive integer and $\lambda_n~\equiv~\frac{\pi}{a}\sqrt{n^2 - \epsilon}$, with $\epsilon\equiv E/E_{\rm t}$. Here, $A_n$ are arbitrary coefficients yet to be determined. Notice that by design, this wave function is continuous across the boundaries separating region F from regions A, B, C, and D. 

To arrive at the expressions in \cref{psiAC,psiBD,psiEsym} we used the fact that the four arms of \cref{fig1} are equivalent and thus the ground state wave function must have even parity.
Details of the general case, where no symmetry is assumed, are sketched in \cref{modeappendix}.

Finally, enforcing continuity of the wave function's derivatives (again, see Appendix~\ref{modeappendix}) results in a matrix equation for the coefficients $A_n$:
\begin{equation}
    M \vec{A} = 0, \label{psiprimeodd}
\end{equation}
where
\begin{equation}
    M_{mn} \equiv  \delta_{mn} \alpha_n \bigl(1 + \tanh \alpha_n \bigr) - \frac{2(2n-1)(2m-1)}{(2n-1)^2 + (2m-1)^2-\epsilon}
\end{equation}
and $\alpha_n \equiv \frac{\pi}{2}\sqrt{(2n-1)^2-\epsilon}$. Here, the vector $\vec{A}$ is a column vector containing the coefficients $A_n$ corresponding to the even-parity basis states. Details of this calculation are provided in \cref{modeappendix}.

Truncating the sums in \cref{psiAC,psiBD,psiEsym} at a cutoff value $N$ results in a finite matrix in \cref{psiprimeodd}, whose solutions exist when $\det M(\epsilon)=0$. In practice, we guess a value of $\epsilon\equiv E/E_{\rm t}$, diagonalize $M(\epsilon)$ for that value of $\epsilon$, and then vary $\epsilon$ until the eigenvalue on the right-hand-side of Eq.~(\ref{psiprimeodd}) becomes zero, as required for a valid solution. As in the matrix mechanics method, the value of $N$ is increased until the results converge to the desired accuracy. In this way, we achieve a ground state energy of $\epsilon=0.659\,606$. This result is accurate to the number of digits quoted.

\begin{figure}[tb]
    \centering
    \includegraphics[width=\linewidth]{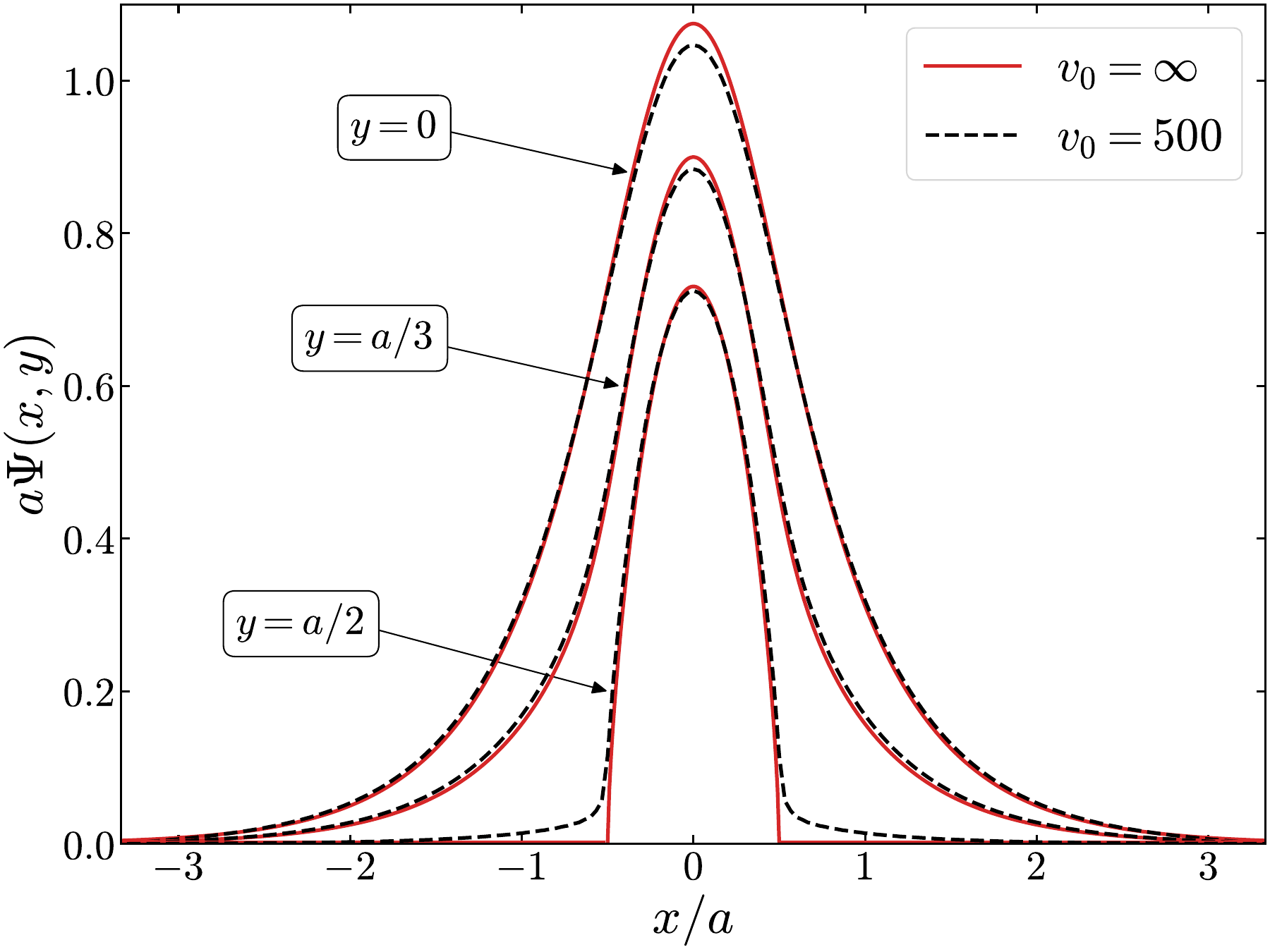}
    \caption{\justifying  
    Ground state wave function for $V_0\rightarrow \infty$ and $L\to\infty$ (solid red curves, obtained via mode matching) and for ${v}_0\equiv V_0/E_{\rm t}=500$ and $a/L = 0.15$ (dashed black curves, obtained via matrix mechanics). Three representative slices of constant $y$ are shown for comparison. For the finite-$V_0$ case, the wave function leaks out of the troughs, which is especially apparent in the $y=a/2$ curve. This makes the wave function less sharply peaked in the centre, most evident in the $y = 0$ slice.}
    \label{psimodes}
\end{figure}

Figure \ref{psimodes} compares the ground state wave function for $V_0\to\infty$ and $L\to\infty$ (computed via mode matching) with the case where $v_0 = V_0/E_{\rm t} = 500$ and $L/a = 20/3$ (computed via matrix mechanics). Because the differences are small, we compare the wave functions along one-dimensional slices at fixed values of $y$. The largest discrepancy between the finite- and infinite-$V_0$ cases can be seen along the edge of the trough at $y=a/2$. For finite $V_0$, an evanescent portion of the wave function exists outside the trough ($|x| > a/2$), unlike the $V_0\rightarrow \infty$ result. This extra weight present in the finite $V_0$ case has to come from somewhere, and therefore the central maximum is correspondingly lower, as seen most clearly in the $y = 0$ slice. In both cases, results are properly converged, and thus the differences are entirely due to the choice of a finite $V_0$; these differences would be accentuated for even smaller values of $V_0$.

One benefit of the mode-matching method is its fast convergence with the truncation value $N$. Truncating the sums at just a single term ($N=1$) yields an energy of $\epsilon=0.6812$, accurate to within approximately $3\%$, while including just 15 terms yields an error of only $10^{-3}$. The $N=1$ case is of particular interest, as it suggests a natural variational wave function which we explore in \cref{sec_var}.

In summary, mode matching is the method best suited to finding the ground state energy of a particle confined to two perpendicular troughs extending to $\pm \infty$ in the $x$- and $y$-directions. We find a bound state pictured as in Fig.~\ref{fig2} with energy $E = 0.659\,606\ E_{\rm t}$, where $E_{\rm t} = \hbar^2 \pi^2/(2m_0a^2)$. On the other hand, mode matching cannot be used for finite barrier potentials.

\section{Tight Binding}\label{tightbinding}

Here, we present an effective model that captures some properties of the present problem in a simplistic way. We use a tight-binding formalism, where a particle hops along two intersecting one-dimensional chains of atoms that meet at a single central site with four neighbours. It is by no means obvious that two intersecting one-dimensional chains would support a bound state, since the trough width has been effectively shrunk to zero and there is no longer a relevant length scale $a$.
To investigate this we write the effective Hamiltonian in second-quantized form as
\begin{equation}
    \hat{H} = -t\sum_{\langle ij\rangle} \Bigl( \hat{c}^\dagger_i\hat{c}^{\vphantom{\dagger}}_j + \hat{c}^\dagger_j\hat{c}^{\vphantom{\dagger}}_i \Bigr), \label{H_TB}
\end{equation}
where the sum is carried out over pairs of nearest-neighbour sites, $t$ is the amplitude for hopping between neighbouring sites, and $\hat{c}^\dagger_i$ ($\hat{c}^{\vphantom{\dagger}}_i$) is the creation (annihilation) operator for a particle at site $i$. 

For a single atomic chain in the thermodynamic limit, the ground state energy is $E_{\rm t,TB}=-2t$, where $E_{\rm t,TB}$ is analogous to the threshold energy in the continuum problem (the subscript $\rm{t}$ stands for ``threshold'', while the additional subscript ${\rm TB}$ stands for tight-binding). For the crossed-chain geometry, we find through a simple matrix diagonalization a ground state energy of $E=-2.3094t<E_{\rm t,TB}$, significantly lower than the minimum energy for the single atomic chain. As was the case with the more complete continuum model, this result indicates that the ground state is a bound state. These calculations are performed numerically for finite crossing chains of $N$ sites each, with periodic boundary conditions, and converged in the thermodynamic limit $N\rightarrow\infty$. In practice, $N\sim30$ is sufficient for convergence.

Interestingly, the tight-binding result turns out to be a low-level approximation to the finite-difference method. As described earlier in Section~\ref{finitedifferences} and in Appendix~\ref{finteappendix}, the finite-difference method requires laying down a set of points in the troughs. Table~\ref{finitegs} in Appendix~\ref{finteappendix} shows results obtained when $N$ points are used to span the width of a trough, $a$; tight-binding corresponds to the choice of $N=1$ (not shown in Table~\ref{finitegs}), with $L \rightarrow \infty$ (also not shown). As described in \cref{finteappendix}, the relationship between $\epsilon$ as defined for the $N=1$ finite-difference method and the energy $E$ as obtained in the tight-binding model is
\begin{equation}
    \epsilon = \frac{4}{\pi^2} \left( \frac{E}{t} + 4\right).
    \label{corresp}
\end{equation}
Using our numerical result, $E/t = -2.3094$ in \cref{corresp} gives $\epsilon = 0.6852$, which is precisely what we obtain using the finite-difference method with $N=1$ and $L \rightarrow \infty$.

In some sense, it is a surprise that the tight-binding approximation yields a bound state, especially since it represents such a crude approximation level for the finite-difference method. Nonetheless, it not only correctly predicts a bound state but also yields a reasonably accurate value for the energy. For this reason, we felt it was worth highlighting the connection between tight binding and the crossed-trough problem.

The success of the crossed one-dimensional tight-binding chains indicates that geometry plays a key role in the creation of a bound state. A more intricate tight-binding model (e.g. one with width in the transverse dimension) would likely capture more closely both qualitative and quantitative aspects of the continuum model depicted in Fig.~\ref{fig1}.

\section{Variational Wave function}\label{sec_var}

It is clear from the previous sections that we have a very good handle on a complete and accurate solution for the bound ground state in the crossed trough geometry. Moreover, the semi-analytic nature of the mode-matching method means it can be used to generate a very simple variational wave function. Truncating the sum in Eq.~(\ref{psiprimeodd}) at $N=1$ results in
\begin{equation}
    \alpha(1 +\tanh\alpha) - \frac{2}{2-\epsilon} = 0, \label{nequals1}
\end{equation}
with $\alpha = \frac{\pi}{2}\sqrt{1 -  \epsilon}$. The solution to this equation is $\epsilon~\approx~0.6812$. The wave function (\cref{psiAC,psiBD,psiEsym}) now becomes very simple:
\begin{align}
    \Psi(x,y) &= A_1\begin{cases}
        e^{-\lambda|x|} \cos \frac{\pi y}{a}, \qquad (x,y) \in \text{ A or C} & \\
        e^{-\lambda|y|} \cos \frac{\pi x}{a}, \qquad (x,y) \in \text{ B or D} & \\
        \frac{e^{-\frac{\lambda a}{2}}}{\cosh\frac{\lambda a}{2}} \Bigl[ \cosh\lambda x \cos \frac{\pi y}{a} + {\cosh\lambda y} \cos \frac{\pi x}{a} \Bigr], & \\
        \phantom{e^{-\lambda|x|} \cos \frac{\pi y}{a}\ ,} \qquad (x,y) \in \text{F} &
    \end{cases} \label{psivar}
\end{align}
where we redefined the normalization constant $A_1$ and where $\lambda$ has become a variational parameter. Unsurprisingly, minimizing the energy expectation value with respect to $\lambda$ (see \cref{variational_appendix} for details) results in a minimum energy of $\epsilon=0.6812$, the same result we achieved from solving \cref{nequals1}. In contrast, the trial wave function given in the textbook problem \cite{griffiths18} yields an energy of $\epsilon = 0.7337$. Our simple variational wave function, motivated by the mode-matching method, outperforms the physically motivated guess given in Ref.~\cite{griffiths18} by a considerable margin, presumably because our choice is the first step in a more systematic scheme that leads to the exact answer. Furthermore, the ground-state wave function achieved by this variational solution closely resembles the exact result, although there are discontinuities in the derivative of the variational wave function at the lines connecting the exterior sections of the troughs to the central region.

\section{Conclusion} \label{summary}

We have used a number of different methods to study the crossed trough problem. Matrix mechanics, finite-difference methods, and mode matching constitute a reasonable toolkit for an undergraduate student in physics. Of these, for this particular problem, we have found mode matching to be the most elegant; it also provides the most accurate solution numerically, and includes the added bonus of providing a natural, simple choice for a variational wave function. The expression for this simple analytical wave function is given in Eq.~(\ref{psivar}) and has an energy of $\epsilon = 0.6812$, quite close to the true ground state energy of $\epsilon = 0.659\,606$.

A variety of extensions are possible for this problem. As geometry plays an important role, one item of interest is where the two troughs are not necessarily perpendicular to one another, along with countless other such variations. Some of these would necessarily require the flexibility of the matrix mechanics method to discuss the impact of rounded edge effects, for example. An extension to three-dimensional tubes would also be interesting. 

\begin{acknowledgments}
This work was supported in part by the Natural Sciences and Engineering Research Council of Canada (NSERC), and the NSERC Alliance -- Alberta Innovates Advance Program.
\end{acknowledgments}
\vskip0.1in
The authors have no conflicts to disclose.

\appendix
\section{Matrix Mechanics Calculation}
\label{matrixappendix}

\begin{figure}[tb]
    \centering
    \includegraphics[scale=0.38]{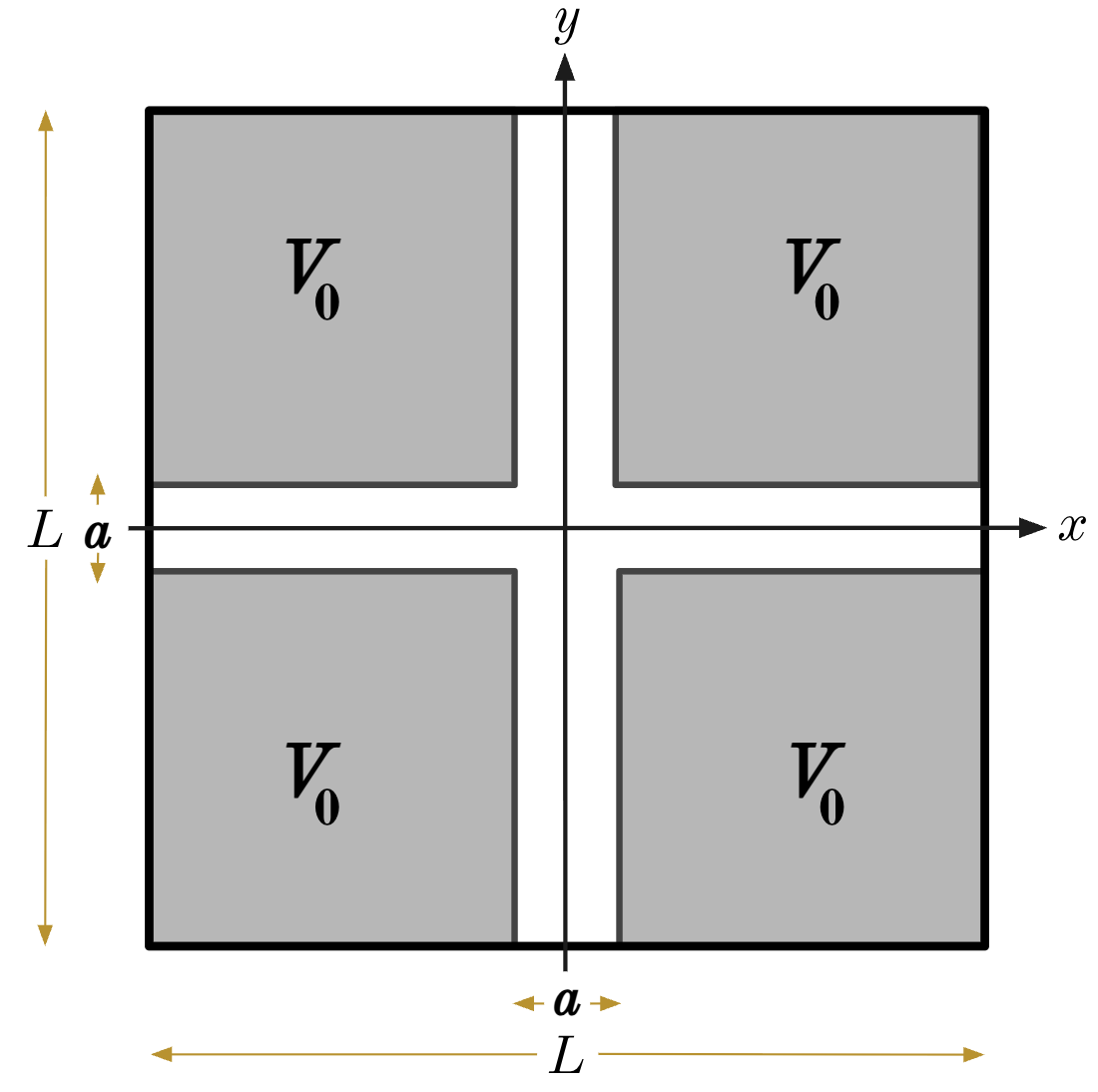}
    \caption{\justifying
    Schematic of the crossed troughs of width $a$ (white regions). The grey-shaded regions have a very large potential $V_0$ that mostly confines the particle to the troughs. The difference from the potential shown in \cref{fig1} is that (i) the entire region shown is now embedded in an infinite two-dimensional square well potential indicated by the outer thick black box -- the potential is infinite outside of this box, and (ii) a finite (i.e. non-infinite) value for $V_0$ is required. For a bound state, $L$ must be chosen large enough so as not to affect the result in any way.}
    \label{appa_fig1}
\end{figure}

As was pointed out in Section~\ref{matrixmechanics}, matrix mechanics requires a basis, which is conveniently arranged by embedding the potential represented in \cref{fig1} in an infinite two-dimensional square well (i.e. a box) as shown in Fig.~\ref{appa_fig1}. The potential representing this confined geometry is given analytically by \cref{V0}. 

For our actual calculations we shift the coordinates $x \rightarrow x^\prime = x + L/2$ and $y \rightarrow y^\prime = y + L/2$, so that the wave function given in Eq.~(\ref{MMexpansion}) uses instead the basis states (we drop the primes)
\begin{equation}
    \phi_{m_x,m_y}(x,y)\equiv \sqrt{\frac{2}{L}} \sin\Bigl(\frac{m_x\pi x}{L}\Bigr) 
    \sqrt{\frac{2}{L}}
    \sin\Bigl(\frac{m_y\pi y}{L}\Bigr), \label{boxbasis_simple}
\end{equation}
and now $0< x,y <L$. These shifted functions retain their evenness or oddness, but now with respect to $x=L/2$ (or $y=L/2$). We write, in bra-ket notation,
\begin{equation}
    H|\psi\rangle = E|\psi\rangle,
    \label{appa3}
\end{equation}
and then expand $|\psi\rangle$ in terms of the basis states
\begin{equation}
    |\psi\rangle = \sum_m c_m|\phi_m\rangle,
    \label{appa4}
\end{equation}
with $|\phi_m\rangle$ the ket representation of the basis state given by Eq.~(\ref{boxbasis_simple}) (and remembering that $m \equiv (m_x,m_y)$ and similarly later for $n$). If we now construct the inner product of both sides of Eq.~(\ref{appa3}) with $\langle \phi_n|$, we arrive at the dimensionless matrix equation (having divided both sides by the energy $E_L \equiv \hbar^2\pi^2/(2m_0L^2)$)
\begin{equation}
    \sum_{m=1}^{N} \bigl[ (h_{0})_{nm} + v_{nm} \bigr]c_m = e c_n,
    \label{a3}
\end{equation}
where $e \equiv E/E_L$,
\begin{equation}
   (h_{0})_{nm} = \delta_{n_xm_x} \delta_{n_ym_y} \left(n_x^2 + n_y^2 \right),
   \label{a4}
\end{equation}
and the matrix element $v_{nm}$ is given by
\begin{equation}
    v_{nm} = {V_0 \over E_L} \left(\frac{b}{L}\right)^2I(n_x,m_x) I(n_y,m_y) P(n_x,m_x,n_y,m_y),
    \label{a5}
\end{equation}
where $b\equiv (L-a)/2$. Here,
\begin{equation}
  \begin{aligned}
    &I(n_j,m_j) = \delta_{n_jm_j} \Bigl[1 - {\rm sinc} \Bigl( \frac{2\pi n_jb}{L} \Bigr) \Bigr] \ \ + \\
    &\bigl[1- \delta_{n_jm_j}\bigr] \biggl[ {\rm sinc} \biggl( \frac{\pi (n_j-m_j)b}{L} \biggr) - {\rm sinc} \biggl(\frac{\pi (n_j+m_j)b}{L} \biggr)  \biggr]
\end{aligned}
\label{a6}
\end{equation}
with $j = x$ or $y$,
\begin{equation}
  \begin{aligned}
    P(n_x,m_x,n_y,m_y) = 1 &+ (-1)^{n_y} (-1)^{m_y} + (-1)^{n_x}(-1)^{m_x} \\
    &+ (-1)^{n_y} (-1)^{m_y}(-1)^{n_x}(-1)^{m_x},
\end{aligned}
\label{a7}
\end{equation}
and
\begin{equation}
    {\rm sinc}(x) \equiv \frac{{\rm sin}(x)}{x}.
    \label{a8}
\end{equation}
To arrive at \cref{mat_eq}, we set $h_{nm}=(h_0)_{nm} + v_{nm}$.

Recall that the two-dimensional box is artificial and should not impact the solutions for bound states. Nonetheless, $E_L$ is the ``natural" energy at this stage (it is the energy of the ground state for a one-dimensional infinite square well of width $L$; the ground state for our two-dimensional empty box is therefore $2E_L$). The length $L$ will drop out of all physical results, provided $L \gg a$. For numerical calculations, the range of the sum in Eq.~(\ref{a3}), in principle infinite, must be truncated at some large cutoff $N$. This is true for \cref{MMexpansion} and (\ref{mat_eq}) as well, so that the matrix $h_{nm}$ becomes of order $N\times N$, where $N = N_{\rm max}\times N_{\rm max}$ and $N_{\rm max}$ is the cutoff for $m_x$ and for $m_y$. The value of $N_{\rm max}$ is increased until the eigenvalues are numerically converged or until computational limitations are reached. Various symmetries can aid in speeding up the computations. For example, the potential is even about $x=L/2$ in $x$ and similarly in $y$, so eigenstates are either even or odd in these variables. If we focus on even eigenstates, which includes the ground state, then only odd integers need to be considered in the expansion \cref{MMexpansion}, and therefore in the matrix equation \cref{a3}. This also implies $P(n_x,m_x,n_y,m_y) = 4$. A similar simplification occurs if one considers only eigenstates that have definite $x\leftrightarrow y$ symmetry.

\begin{figure}[tb]
    \centering
    \includegraphics[width=1.0\linewidth]{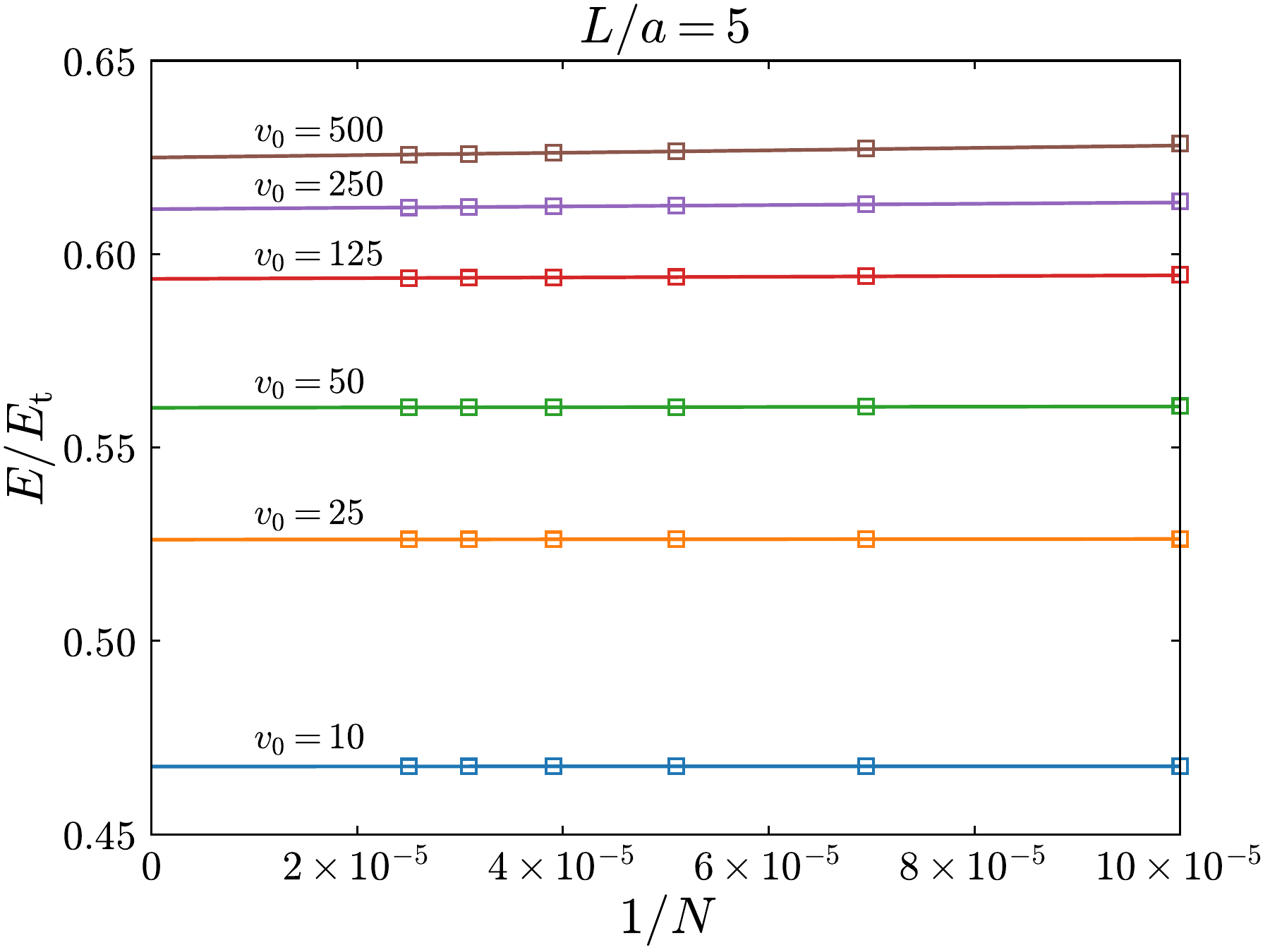}
    \caption{\justifying
    Normalized energies $\epsilon \equiv E/E_{\rm t}$ versus $1/N$ for various values of ${v}_0$, for $L/a=5$. Matrices of size $N\times N$ ($N= 10\,000,\ 14\,400,\ 19\,600,\ 25\,600,\ 32\,400$, and $40\,000$) were used to obtain the results indicated with square symbols. The straight lines were determined by manual fits for each value of ${v}_0$. Fit lines are given (in order of increasing value of ${v}_0$) by 
    $\epsilon = 0.46753 + 0.51/N$, $\epsilon = 0.52619 + 1.58/N$, $\epsilon = 0.56027 + 3.3/N$, $\epsilon = 0.59359 + 9.0/N$, $\epsilon = 0.61162 + 17.2/N$, and $\epsilon = 0.62495 + 31.0/N$. The precision of these fits is indicated by the number of significant digits, particularly in the intercepts. The $N \rightarrow \infty$ extrapolated values are used in \cref{fig3}.}
    \label{figa1_1}
\end{figure}

In Fig.~\ref{figa1_1} we show the normalized ground state energy $\epsilon \equiv E/E_{\rm t}$ versus $1/N$ for a variety of barrier strengths ${v}_0 \equiv V_0/E_{\rm t}$, for the case $L/a=5$.
Here, $N$ denotes the matrix size before accounting for any symmetries.
On the scale of this plot, the energy for each value of $v_0$ appears nearly constant as a function of $N$. There is, however, a small dependence that becomes more pronounced as $v_0$ increases. This dependence is approximately linear in $1/N$ for large values of $N$; for example, one can plot the 
$v_0=125$ results and determine that $\epsilon = 0.59359 + 9.0/N$ is a good fit, particularly to the lowest 3 points. The intercepts of the fit lines, as tabulated in the second column of Table~\ref{table_a1}, give an estimate for the exact value of the energy in the $N\to\infty$ limit. 
A conservative estimate for the error in intercept values is $\pm 0.000\,05$.

These results were obtained using a program written in Fortran. The numerical diagonalization was achieved by calling the subroutine
\texttt{dsyevd.f}. Various results were benchmarked by two independent codes (written by other coauthors) in Python. Similar redundancy and benchmarking exists for other codes used for the mode-matching, finite-difference, and tight-binding calculations.

Table~\ref{table_a1} also contains values of the ground state energy for a single trough with barrier height $V_0$. The calculation for this energy follows that of a single one-dimensional square well (see, e.g. Ref.~[\onlinecite{griffiths18}], section 2.6), and requires solving the transcendental equation
\begin{equation}
    \tan(z) = \sqrt{\biggl({\frac{z_0}{z}}\biggr)^2 - 1},
\label{trans}
\end{equation}
where $z~\equiv~\frac{\pi}{2} \sqrt{{E_{\rm t}^{V_0}}/{E_{\rm t}}}$ and $z_0~\equiv~\frac{\pi}{2}\sqrt{{ v}_0}$, with $E_{\rm t}^{V_0}$ the ground state energy for this problem. A solution for $E_{\rm t}^{V_0}$ exists no matter how small $V_0$ is, but this does not automatically imply that a bound state exists in the geometry of the crossed trough (Fig.~\ref{fig1}) for arbitrarily small $V_0$. The inset in Fig.~\ref{fig3} illustrates how $E_{\rm t}^{V_0}$ varies with ${v}_0$.

\begin{table}[t]
    \centering
    \begin{tabular}{|c|c|c|}
        \hline
        $v_0$ & $\epsilon$ & $E_{\rm t}^{V_0}/E_{\rm t}$ \\
        \hline
        \phantom{aaaa} 10 \phantom{aaaa}  & \phantom{aaaa} 0.46753 \phantom{aaaa} & \phantom{aaaa} 0.6901730 \phantom{aaaa} \\
        25 & 0.52619 & 0.7859253 \\
        50 & 0.56027 & 0.8412384 \\
        125 & 0.59359 & 0.8950400 \\
        250 & 0.61162 & 0.9240440 \\
        500 & 0.62495 & 0.9453851 \\
        \hline
    \end{tabular}
    \caption{\justifying 
    Values of $\epsilon \equiv E/E_{\rm t}$ for ${\rm v}_0 \equiv V_0/E_{\rm t}$, as extrapolated from linear fits to the data in Fig.~\ref{figa1_1}, for the case $L/a = 5$. The estimated error is $\pm 0.000\,05$. These values of $\epsilon$ are plotted in Fig.~\ref{fig3} (lower points in the main figure). The third column contains the normalized threshold value $E_{\rm t}^{V_0}/E_{\rm t}$, as determined below and plotted in the inset of Fig.~\ref{fig3}.}
\label{table_a1}
\end{table}

One can expand Eq.~(\ref{trans}) around the ${v}_0 \rightarrow \infty$ limit, and obtain
\begin{equation}
    \frac{E_{\rm t}^{V_0}}{E_{\rm t}} =  1 - \frac{4}{\pi} \frac{1}{\sqrt{{v}_0}} + 3\left(\frac{2}{\pi}\right)^2\frac{1}{{v}_0} +\ \cdots.
    \label{expansion}
\end{equation}
The dashed black curve in the inset of Fig.~\ref{fig3}
represents the first-order
correction to unity. Including further terms in this expansion makes the result essentially indistinguishable from the numerical solution to Eq.~(\ref{trans}) on the scale of this inset.

\section{Finite-Difference Calculation}\label{finteappendix}

\begin{figure}[tb]
    \centering
    \includegraphics[width=\linewidth]{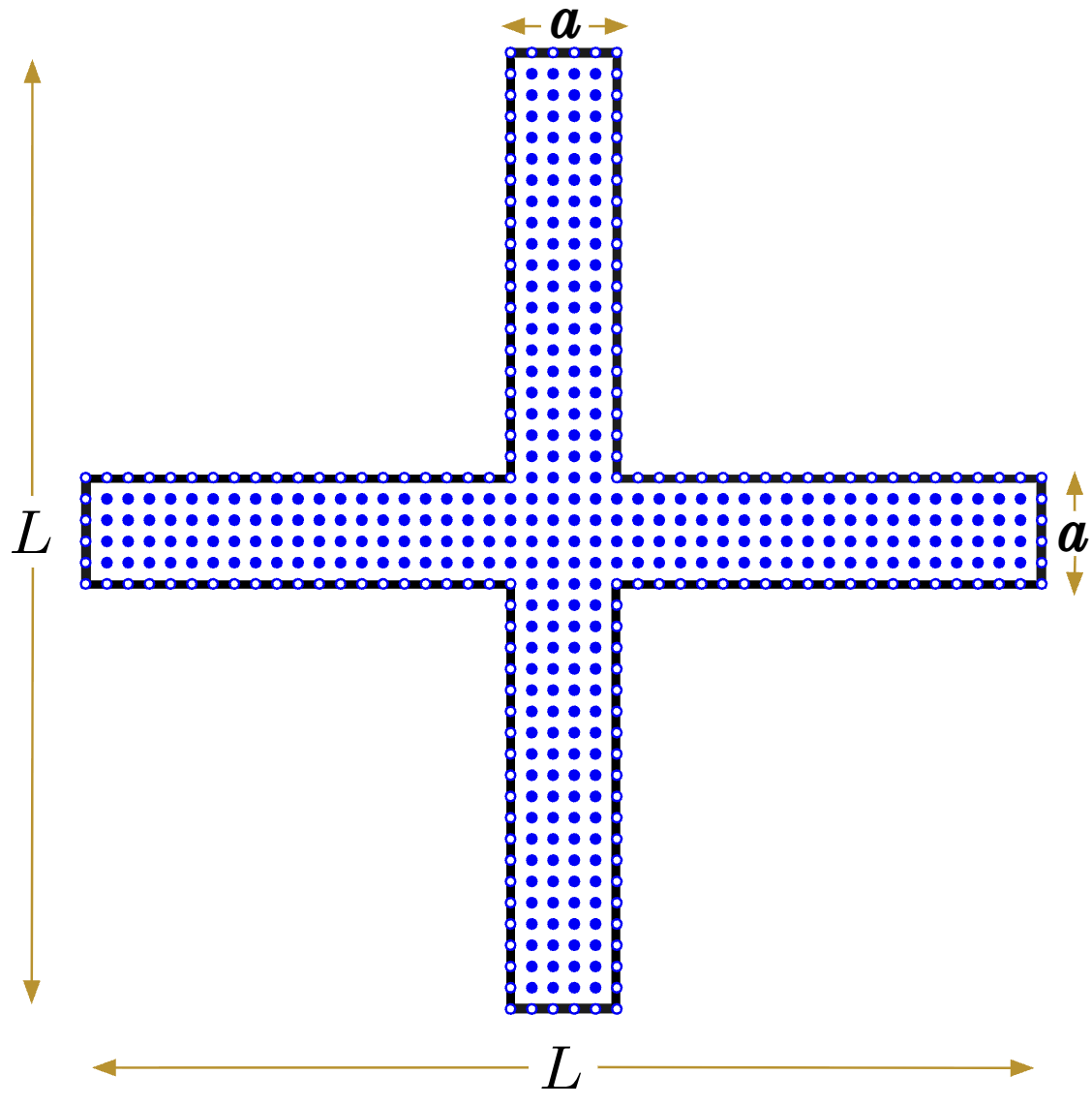}
    \caption{\justifying 
    Grid of points used to discretize the geometry of the problem. Points are separated by a distance $h$ in both the $x$- and $y$-directions such that $N$ interior points span the width of each trough (here $N=4$). Each interior point (filled circle) is associated with a variable representing the value of the wave function at that point. Points on the boundary (unfilled circles) carry a value of $0$ to satisfy the boundary conditions, and are thus only included implicitly.}
    \label{figpoints}
\end{figure}

To solve the Schr{\"o}dinger equation (\ref{SE_no_V}) using finite differences, once more the potential must be truncated at $x,y = \pm L/2$. In this case we take the potential $V_0$ to be infinite, so there is no distinction between the shaded grey regions in \cref{appa_fig1} and the region outside the bounding box.  The region where $V=0$ is discretized by introducing a grid of points, separated by a distance $h=a/(N+1)$, where $N$ is the number of interior grid points spanning the width of each trough, as shown in \cref{figpoints}. To each of the $M$ interior points $(x,y)$ we associate a variable $\Psi_{x,y}$ representing the value of the wave function at that point.

We substitute the discrete Laplacian \cite{abramowitz64}
\begin{equation}
    \nabla^2\Psi(x,y) \approx \frac{\Psi_{x_+,y}+ \Psi_{x_-,y}+\Psi_{x,y_+}+\Psi_{x,y_-}-4\Psi_{x,y}}{h^2}, \label{laplacian}
\end{equation}
where $x_\pm\equiv~x\pm~h$ and similarly for $y_\pm$, into \cref{SE_no_V} to obtain an equation of the form
\begin{equation}
    \biggl(\frac{N+1}{\pi}\biggr)^2 \Bigl[4\Psi_{x,y} - \Psi_{x_+,y} - \Psi_{x_-,y} - \Psi_{x,y_+} - \Psi_{x,y_-}\Bigr] = \epsilon\Psi_{x,y}, \label{SE_finite}
\end{equation}
with $\epsilon\equiv E/E_{\rm t}$, for each of the $M$ variables $\Psi_{x,y}$. Together, Eqs.~(\ref{SE_finite}) define rows of the $M\times M$ matrix \cref{Laplacian_matrix}, where we set $\Psi_{x,y}=0$ for points along the boundary as necessary, with
\begin{equation}
    \bar{L}_{ij} = \biggl(\frac{N+1}{a}\biggr)^2 \times \begin{cases}
        \phantom{-}4 & \text{if }i=j, \\
        -1 & \text{if }i\text{  
        is next to }j, \\
        \phantom{-}0 & \text{otherwise,}
    \end{cases}
\end{equation}
and $i,j\equiv(x,y)$. Equation~(\ref{Laplacian_matrix}) is then solved by numerical diagonalization, and the resulting eigenvalues are tabulated in Table~\ref{finitegs}.

\begin{table}[tb]
    \begin{tabular}{|c|c c c c c c |}
    \hline
        \backslashbox{$L/a$}{$N$} & 50 & 100 & 200 & 250 & 300 & 350 \\
        \hline
        5  & 0.66166 & 0.66063 & 0.66020 & 0.66013 & 0.66008 & 0.66005 \\
        7  & 0.66132 & 0.66032 & 0.65990 & 0.65983 & 0.65978 & 0.65975 \\
        9  & 0.66131 & 0.66031 & 0.65989 & 0.65982 & 0.65978 & 0.65974 \\
        \hline
    \end{tabular}
    \caption{\justifying 
    Ground state energy $\epsilon\equiv E/E_{\rm t}$, found numerically via finite differences. Rows correspond to a given arm length  $L$ in terms of the trough width $a$. Columns correspond to the number of points $N$ used to span each trough. Points on the boundary are not included, so the mesh spacing is $h=a/(N+1)$. As is clear from examining the second and third rows, making the arms three times as long as they are wide ($L=7a$) is sufficient to converge the results to $4$ decimal places. 
    For comparison, our most accurate result is $\epsilon = 0.659\,606$, found with the mode-matching method.}
    \label{finitegs}
\end{table}

To make the connection between finite differences and tight binding described in \cref{tightbinding}, we substitute $N=1$ into \cref{SE_finite} and move the diagonal term to the right-hand side to arrive at
\begin{equation}
     -\Bigl( \Psi_{x_+,y} + \Psi_{x_-,y} + \Psi_{x,y_+} + \Psi_{x,y_-}\Bigr) = \biggl( \frac{\pi^2}{4}\epsilon - 4\biggr) \Psi_{x,y}. \label{SE_N1}
\end{equation}
The relation in \cref{SE_N1} can be understood as a matrix representation of the Schr{\"o}dinger equation $H\Psi = E\Psi$, using the Hamiltonian from \cref{H_TB}, where we identify
\begin{equation}
    \biggl( \frac{\pi^2}{4}\epsilon - 4\biggr) \equiv E/t. \label{FD_TB_connection2}
\end{equation}
Solving \cref{FD_TB_connection2} for $\epsilon$ gives \cref{corresp}.

\section{Mode-Matching Calculation}\label{modeappendix}

Solving the Schr{\"o}dinger equation (\ref{SE_no_V})
on the regions of \cref{fig1} is done by separation of variables, a process familiar to students at the undergraduate level. Assuming a separable solution of the form $\Psi(x,y) = X(x)Y(y)$, \cref{SE_no_V} becomes
\begin{equation}
\begin{aligned}
    Y''(y) &= -\phi^2 Y(y), \\
    X''(x) &=  \lambda^2 X(x),
\end{aligned}
\end{equation}
for constants $\phi$ and $\lambda = \sqrt{\phi^2-{2m_0E}/{\hbar^2}}$.

Consider region A for definiteness. For $\Psi$ to vanish on both edges of the trough, $Y(y)$ must take the form
\begin{equation}
    Y_n(y) = \sin \Bigl(\phi_n\Bigl(y + \frac{a}{2}\Bigr)\Bigr), \quad n = 1,2,\dots,
\end{equation}
with $\phi_n=n\pi/a$. For a bound state to exist, we need $E<E_{\rm t}$ which implies $\lambda^2>0$, and therefore
\begin{equation}
    X_n(x) = e^{\lambda_n\bigl(\frac{a}{2}+x\bigr)},
\end{equation}
where now $\phi$ and $\lambda$ have acquired a dependence on the quantum number $n$, and we have used only the one solution that remains finite as $x \rightarrow -\infty$. The general solution on region A is the linear combination
\begin{equation}
    \Psi_{\rm A}(x,y) = \sum_{n=1}^\infty A_n e^{\lambda_n\bigl(\frac{a}{2}+x\bigr)} \sin \biggl(\frac{n\pi}{a}\Bigl(y + \frac{a}{2}\Bigr)\biggr),
    \label{appc_regiona}
\end{equation}
with arbitrary coefficients $A_n$. Similarly, for the other three trough regions we have 
 \begin{align}
      \Psi_{\rm B}(x,y) &= \sum_{n=1}^\infty B_n e^{\lambda_n\bigl(\frac{a}{2}-y\bigr)} \sin \biggl(\frac{n\pi}{a}\biggl(x + \frac{a}{2}\biggr)\biggr), \\
     \Psi_{\rm C}(x,y) &= \sum_{n=1}^\infty C_n e^{\lambda_n\bigl(\frac{a}{2}-x\bigr)} \sin  \biggl(\frac{n\pi}{a}\biggl(y + \frac{a}{2}\biggr)\biggr), \\
     \Psi_{\rm D}(x,y) &= \sum_{n=1}^\infty D_n e^{\lambda_n\bigl(\frac{a}{2}-y\bigr)} \sin \biggl(\frac{n\pi}{a}\biggl(x + \frac{a}{2}\biggr)\biggr).
     \label{psi}
 \end{align}

Region F has no external boundaries at which to impose boundary conditions, so we consider four separate boundary-value problems such that
\begin{equation}
    \Psi_{\rm F} =  \Psi_{\rm F_{\rm A}} + \Psi_{\rm F_{\rm B}} + \Psi_{\rm F_{\rm C}} + \Psi_{\rm F_{\rm D}},
\end{equation}
where $\Psi_{\rm F_{\rm A}}$ vanishes on all boundaries of F except its border with region A; $\Psi_{\rm F_{\rm B}}$ is nonzero only on the border with B; and so on. Again, the condition $\lambda_n^2>0$ leads to
\begin{equation}
\begin{aligned}
    \Psi_{\rm F}(x,y) &= \sum_{n=1}^\infty  A_n' \sinh \Bigl(\lambda_n \Bigl( \frac{a}{2} - x\Bigr)\Bigr) \sin\biggl( \frac{n\pi}{a}\Bigl(y + \frac{a}{2}\Bigr)\biggr) \\
    &\,+ \sum_{n=1}^\infty B_n' \sinh \Bigl(\lambda_n \Bigl( \frac{a}{2} + y\Bigr)\Bigr) \sin\biggl( \frac{n\pi}{a}\Bigl(x + \frac{a}{2}\Bigr)\biggr) \\ 
    &\,+ \sum_{n=1}^\infty C_n' \sinh \Bigl(\lambda_n \Bigl( \frac{a}{2} + x\Bigr)\Bigr) \sin\biggl( \frac{n\pi}{a}\Bigl(y + \frac{a}{2}\Bigr)\biggr) \\ 
    &\,+ \sum_{n=1}^\infty D_n' \sinh \Bigl(\lambda_n \Bigl( \frac{a}{2} - y\Bigr)\Bigr) \sin\biggl( \frac{n\pi}{a}\Bigl(x + \frac{a}{2}\Bigr)\biggr).
\end{aligned}
\label{appc_primed}
\end{equation}
Matching the wave function along the dashed boundary lines in \cref{fig1} gives a simple relation between the primed coefficients in Eq.~(\ref{appc_primed}) and their un-primed counterparts in the corresponding troughs. For example, concerning the term in F that shares a boundary with region A, we have $A_n'=A_n/\sinh (\lambda_na)$, and similarly for the other coefficients.

The wave function's derivative must also be continuous across inter-region boundaries. We take the inner product of this condition with appropriate sine functions to arrive at
\begin{equation}
\begin{aligned}
    \int_{-\frac{a}{2}}^{\frac{a}{2}} dy\,\sin \biggl(\frac{m\pi}{a}\Bigl(y+\frac{a}{2} \Bigr)\biggr) \biggl[\frac{\partial\Psi_{\rm A}}{\partial x} - \frac{\partial\Psi_{\rm F}}{\partial x} \Biggr]_{x=-\frac{a}{2}}  &= 0, \\
    \int_{-\frac{a}{2}}^{\frac{a}{2}} dx\,\sin \biggl(\frac{m\pi}{a}\Bigl(x+\frac{a}{2} \Bigr)\biggr) \biggl[\frac{\partial\Psi_{\rm B}}{\partial y} - \frac{\partial\Psi_{\rm F}}{\partial y}  \Biggr]_{y=\frac{a}{2}\phantom{-}}  &= 0, \\
    \int_{-\frac{a}{2}}^{\frac{a}{2}} dy\,\sin \biggl(\frac{m\pi}{a}\Bigl(y+\frac{a}{2} \Bigr)\biggr) \biggl[ \frac{\partial\Psi_{\rm C}}{\partial x} - \frac{\partial\Psi_{\rm F}}{\partial x} \Biggr]_{x=\frac{a}{2}\phantom{-}}  &= 0, \\
    \int_{-\frac{a}{2}}^{\frac{a}{2}} dx\,\sin \biggl(\frac{m\pi}{a}\Bigl(x+\frac{a}{2} \Bigr)\biggr) \biggl[ \frac{\partial\Psi_{\rm D}}{\partial y} - \frac{\partial\Psi_{\rm F}}{\partial y}   \Biggr]_{y=-\frac{a}{2}}  &= 0. \label{continuous_derivative}
\end{aligned}
\end{equation}
For the case of fourfold rotational symmetry, the equations in \cref{continuous_derivative} are all equivalent; performing the derivatives in the first equation, for example, gives
\begin{equation}
    A_m \frac{\lambda_ma}{2}\Bigl(1 + \tanh \frac{\lambda_ma}{2}\Bigr) - \sum_{n=1}^\infty A_n \rho_{mn} = 0
\end{equation}
for all integers $m\geq1$, where
\begin{equation}
\begin{aligned}
    \rho_{mn} &\equiv \frac{1-(-1)^m}{\sinh\lambda_na}\frac{n\pi}{a} \\
    &\qquad\times \int_{-\frac{a}{2}}^{\frac{a}{2}} dy \sinh \biggl(\lambda_n\Bigl(\frac{a}{2}+y\Bigr)\biggr)\sin \biggl( \frac{m\pi}{a}\Bigl(y+\frac{a}{2}\Bigr)\biggr) \\
    &= \frac{1-(-1)^m}{2}\frac{2nm}{n^2 + m^2-\epsilon},
\end{aligned}    
\end{equation}
with $\epsilon\equiv E/E_{\rm t}$. Putting this in the form of a matrix equation yields
\begin{equation}
    \bigl(D - \rho\bigr) \vec{A} = 0, \label{psiprime}
\end{equation}
with
\begin{align}
    D_{mn} &\equiv \delta_{nm} \frac{\lambda_na}{2} \Bigl( 1 + \tanh \frac{\lambda_na}{2} \Bigr), \\
    \rho_{mn} &\equiv \frac{1-(-1)^m}{2} \frac{2nm}{m^2 + n^2 - \epsilon}. 
    \label{mat_prelim}
\end{align}

Finally, symmetry requires that $A_n=0$ for all even $n$, since these modes have odd symmetry with respect to reflections across the coordinate axes. For this reason, all even rows of the matrix can be removed, and \cref{psiprime} can be written more compactly in a symmetric form as \cref{psiprimeodd}.

\section{Variational Wave Function}\label{variational_appendix}

The wave function in \cref{psivar} can be normalized analytically, though the algebra is somewhat tedious. The arms are symmetric so we need only integrate over one of them (say, region C in Fig.~\ref{fig1}):
\begin{equation}
    \int_{\rm F} \bigl|\Psi_{\rm F}(x,y)\bigr|^2 + 4\int_{\rm C} \bigl|\Psi_{\rm C}(x,y)\bigr|^2 \stackrel{!}{=} 1. \label{norm_conidition}
\end{equation}
The integrals here are all elementary, and therefore from \cref{norm_conidition} we have
\begin{equation}
    A_1^2 = \frac{e^{\lambda a}}{a^2} \Biggl[ \frac{1}{\lambda a}\Bigl(1 + \tanh \frac{\lambda a}{2} \Bigr) + \frac{1}{2}\sech^2 \frac{\lambda a}{2} + \frac{8\pi^2}{\bigl(\pi^2 + a^2\lambda^2\bigr)^2}\Biggr]^{-1}.
\end{equation}

We next consider the expectation value of the Hamiltonian. Because the variational wave function (unlike the infinite sum in \cref{psi}) is discontinuous across region boundaries, the integral must be explicitly evaluated across these boundaries, as well as on the interior of regions A through F. Again using symmetry,
\begin{equation}
\begin{aligned}
    \braket{\hat{H}} &= -\frac{\hbar^2}{2m_0} \Biggl[ \int_{\rm F} \Psi_{\rm F}(x,y) \nabla^2\Psi_{\rm F}(x,y) \\
    &\qquad\qquad+ 4\int_{\rm C} \Psi_{\rm C}(x,y) \nabla^2\Psi_{\rm C}(x,y) \\
    &\qquad\qquad+ 4\lim_{\epsilon\rightarrow0} \int_{\frac{a}{2}-\epsilon}^{\frac{a}{2}+\epsilon} dx\,\int_{-\frac{a}{2}}^{\frac{a}{2}}dy\, \Psi(x,y) \nabla^2\Psi(x,y) \Biggr]. \label{H_expectation}
\end{aligned}
\end{equation}
\begin{widetext}
The integrals over individual regions are evaluated easily by exploiting the fact that, by construction, the wave functions are eigenfunctions of the Laplacian, so we have
\begin{equation}
    \int_{\rm F} \Psi_{\rm F}(x,y) \nabla^2\Psi_{\rm F}(x,y) + 4\int_{\rm C} \Psi_{\rm C}(x,y) \nabla^2\Psi_{\rm C}(x,y) = \biggl[\lambda^2-\frac{\pi^2}{a^2} \biggr] \underbrace{\biggr[ \int_{\rm F} \bigl|\Psi_{\rm F}(x,y)\bigr|^2 + 4\int_{\rm C} \bigl|\Psi_{\rm C}(x,y)\bigr|^2 \biggr]}_{=1}.
\end{equation}
For the boundary integral, we write the wave function in the vicinity of the boundary between regions C and F as
\begin{equation}
\begin{aligned}
    \Psi(x,y) &= A_1 \biggl[ e^{-\lambda x} \cos \frac{\pi y}{a} \theta\Bigl(x-\frac{a}{2}\Bigr)
    + \frac{e^{-\frac{\lambda a}{2}}}{\cosh \frac{\lambda a}{2}} \Bigl[ \cosh\lambda x \cos \frac{\pi y}{a} + {\cosh\lambda y} \cos \frac{\pi x}{a} \Bigr] \theta\Bigl(\frac{a}{2}-x\Bigr)\biggr].
\end{aligned}
\end{equation}
Then
\begin{equation}
\begin{aligned}
    \frac{\partial\Psi}{\partial x} &= A_1 \Biggl\{ -\lambda e^{-\lambda x} \cos \frac{\pi y}{a} \theta\Bigl(x-\frac{a}{2}\Bigr)
    + \frac{e^{-\frac{\lambda a}{2}}}{\cosh \frac{\lambda a}{2}} \Bigl[\lambda \sinh\lambda x \cos \frac{\pi y}{a} -\frac{\pi}{a} {\cosh\lambda y} \sin \frac{\pi x}{a} \Bigr] \theta\Bigl(\frac{a}{2}-x\Bigr) \\
    & \qquad +\biggl( \biggl[ e^{-\lambda x} -\frac{e^{-\frac{\lambda a}{2}}}{\cosh \frac{\lambda a}{2}} \cosh\lambda x \biggr] \cos \frac{\pi y}{a}
    - \frac{e^{-\frac{\lambda a}{2}}}{\cosh \frac{\lambda a}{2}} \cosh\lambda y \cos\frac{\pi x}{a}  \biggr) \delta\Bigl(x-\frac{a}{2}\Bigr)\Biggr\}.
\end{aligned}
\end{equation}
The last term (second line proportional to the $\delta$-function) can be dropped because the quantity in parentheses vanishes at $x=a/2$. Indeed, any function $f(x)$ for which $f(x_0)=0$ satisfies $f(x_0)\delta(x-x_0) \equiv 0$ even after taking further derivatives, since for arbitrary $g(x)$ we have
\begin{equation}
\begin{aligned}
    \phantom{a} & \int dx\, g \frac{d}{dx} \Bigl( f\delta(x-x_0) \Bigr)
    = \int dx\, \biggl( gf'\delta(x-x_0) + gf\frac{d}{dx}\delta(x-x_0)\biggr) \\
    =& g(x_0)f'(x_0) + \cancelto{0}{g(x)f(x)\delta(x-x_0)} \ - \ \int dx\, (gf)' \delta(x-x_0)
    = g(x_0)f'(x_0) - \cancelto{0}{g'(x_0)f(x_0)} - g(x_0)f'(x_0) = 0.
\end{aligned}
\end{equation}
Taking another derivative,
\begin{equation}
\begin{aligned}
    \frac{\partial^2\Psi}{\partial^2 x} &= A_1 \Biggl\{ \lambda^2 e^{-\lambda x} \cos \frac{\pi y}{a} \theta\Bigl(x-\frac{a}{2}\Bigr)
    + \frac{e^{-\frac{\lambda a}{2}}}{\cosh \frac{\lambda a}{2}} \Bigl[\lambda^2 \cosh\lambda x \cos \frac{\pi y}{a}
    - \biggl(\frac{\pi}{a}\biggr)^2 {\cosh\lambda y} \cos \frac{\pi x}{a} \Bigr] \theta\Bigl(\frac{a}{2}-x\Bigr) \\
    &\qquad - \biggl(\biggl[ \lambda e^{-\lambda x}  + \frac{e^{-\frac{\lambda a}{2}}}{\cosh \frac{\lambda a}{2}} \lambda \sinh\lambda x \biggr] \cos \frac{\pi y}{a}
    - \frac{\pi}{a} \frac{e^{-\frac{\lambda a}{2}}}{\cosh \frac{\lambda a}{2}} {\cosh\lambda y} \sin \frac{\pi x}{a} \biggr) \delta\Bigl(x-\frac{a}{2}\Bigr) \Biggr\}. \label{x_derivative}
\end{aligned}
\end{equation}
Since we take the range of integration around $x=a/2$ to be $2\epsilon\to 0$, only the $\delta$-function terms in \cref{x_derivative} will survive in the final result. For the same reason, all terms coming from the $y$-derivative vanish in the limit. The remaining integral is
\begin{equation}
\begin{aligned}
    \phantom{=}& \int_{\frac{a}{2}-\epsilon}^{\frac{a}{2}+\epsilon} dx\, \int_{-\frac{a}{2}}^{\frac{a}{2}} dy\, \Psi(x,y) \nabla^2\Psi(x,y) \\
    =& A_1 \int_{\frac{a}{2}-\epsilon}^{\frac{a}{2}+\epsilon} dx\, \int_{-\frac{a}{2}}^{\frac{a}{2}} dy\, \Psi(x,y) \biggl( - \biggl[ \lambda e^{-\lambda x}  + \frac{e^{-\frac{\lambda a}{2}}}{\cosh \frac{\lambda a}{2}} \lambda \sinh\lambda x \biggr]  \cos\frac{\pi y}{a}
    -\frac{\pi}{a} \frac{e^{-\frac{\lambda a}{2}}}{\cosh \frac{\lambda a}{2}} {\cosh\lambda y} \sin \frac{\pi x}{a} \biggr) \delta\Bigl(x-\frac{a}{2}\Bigr) \\
    =& A_1 \int_{-\frac{a}{2}}^{\frac{a}{2}} dy\, {\Psi\Bigl(\frac{a}{2},y\Bigr)} \biggl( -\lambda e^{-\frac{\lambda a}{2}} \biggl[ 1 +\tanh \frac{\lambda a}{2} \biggr] \cos \frac{\pi y}{a}
    -\frac{\pi}{a} \frac{e^{-\frac{\lambda a}{2}}}{\cosh \frac{\lambda a}{2}} \cosh\lambda y \biggr) \\
    =& A_1^2 \Biggl[ -\frac{\lambda a}{2} \Bigl(1  +\tanh\frac{\lambda a}{2}\Bigr) + \frac{2\pi^2 }{\pi^2 + \lambda^2a^2} \Biggr].
\end{aligned}
\end{equation}
All together, \cref{H_expectation} becomes
\begin{equation}
\begin{aligned}
    \braket{\hat{H}} &= \frac{\hbar^2\pi^2}{2m_0a^2} \Biggl[1- \frac{\lambda^2a^2}{\pi^2}
    + \frac{4}{\pi^2}\frac{ \frac{\lambda a}{2} \Bigl(1  +\tanh\frac{\lambda a}{2}\Bigr) - \frac{2\pi^2 }{\pi^2 + \lambda^2a^2}}{\frac{1}{\lambda a}\Bigl(1 + \tanh \frac{\lambda a}{2} \Bigr) + \frac{1}{2}\sech^2 \frac{\lambda a}{2} + \frac{8\pi^2}{\bigl(\pi^2 + a^2\lambda^2\bigr)^2}}  \Biggr].
\end{aligned}
\label{appd10}
\end{equation}
The minimum value of this expression can be determined using a numerical minimization algorithm to be $ \braket{\hat{H}}_{\rm min} = 0.6812\,E_{\rm t}$, achieved when $\lambda~=~1.7738/a$; this is precisely the result achieved via \cref{nequals1}. Alternatively, taking the derivative of Eq.~(\ref{appd10}) with respect to $\lambda$, and setting the result to zero, one finds, after much algebra, the result
\begin{equation}
\lambda a \Bigl( 1 + \tanh\frac{\lambda a}{2}\Bigr)  =\frac{4}{1 + \bigl( \frac{\lambda a}{\pi}\bigr)^2},
\label{appd11}
\end{equation}
which, taken with Eq.~(\ref{appd10}), is precisely \cref{nequals1} for $\alpha = \lambda a/2$.
\end{widetext}

\bibliography{refs}
\end{document}